\documentclass[12pt]{article}
\begin{document}
\begin{center}
{\large\bf Positive and Negative Energy Symmetry and the
Cosmological Constant Problem} \vskip 0.3 true in {\large J. W.
Moffat} \vskip 0.3 true in {\it The Perimeter Institute for
Theoretical Physics, Waterloo, Ontario, N2L 2Y5, Canada} \vskip
0.3 true in and \vskip 0.3 true in {\it Department of Physics,
University of Waterloo, Waterloo, Ontario N2L 3G1, Canada}
\end{center}
%\date{\today}
\begin{abstract}%
The action for gravity and the standard model includes, as well as
the positive energy fermion and boson fields, negative energy
fields. The Hamiltonian for the action leads through a positive
and negative energy symmetry of the vacuum to a cancellation of
the zero-point vacuum energy and a vanishing cosmological constant
in the presence of a gravitational field solving the cosmological
constant problem. To guarantee the quasi-stability of the vacuum,
we postulate a positive energy sector and a negative energy sector
in the universe which are identical copies of the standard model.
They interact only weakly through gravity. As in the case of
antimatter, the negative energy matter is not found naturally on
Earth or in the universe. A positive energy spectrum and a
consistent unitary field theory for a pseudo-Hermitian Hamiltonian
is obtained by demanding that the pseudo-Hamiltonian is ${\cal
P}{\cal T}$ symmetric. The quadratic divergences in the two-point
vacuum fluctuations and the self-energy of a scalar field are
removed. The finite scalar field self-energy can avoid the Higgs
hierarchy problem in the standard model.
\end{abstract} \vskip 0.2 true in e-mail:
john.moffat@utoronto.ca

%\pacs{ }

\section{Introduction}

The cosmological constant problem is considered to be one of the
major problems of modern physics~\cite{Weinberg}. The particle
physics origin of the problem arises because of the quartic
divergence of the zero-point vacuum energy in the presence of a
gravitational field. The constant zero-point energy cannot be
shifted to zero in the action due to the universal coupling of
gravity to energy including vacuum energy. It is hoped that a
natural symmetry exists that explains why the cosmological
constant is zero or small. However, the obvious candidates,
supersymmetry and conformal invariance, cannot supply this
solution, for they are badly broken in Nature. We shall introduce
a positive and negative energy symmetry of particles that removes
the generic infinity of the zero-point vacuum energy in the
presence of a gravitational field.

Kaplan and Sundrum~\cite{Sundrum} have introduced a a discrete
parity symmetry which transforms positive energy into negative
energy through a projection operator $P$. As in the case of
Linde's~\cite{Linde,Linde2} two--universe proposal, the authors
postulate that to avoid a breakdown of the vacuum due to the
negative energy particles, the two copies of the standard model
matter fields, corresponding to positive and negative energy
particles, interact only weakly through gravity. To prevent
excessively rapid decay of the vacuum, it is also postulated that
gravitational Lorentz invariance breaks down at short distances.
't Hooft and Nobbenhuiz~\cite{Nobbenhuiz2} have studied a symmetry
argument exploiting a complex space with both positive and
negative energy particles.

A local quantum field theory with positive and negative energy
particle symmetry was recently developed~\cite{Moffat}, following
a method of quantizing fields introduced by Dirac in
1942~\cite{Dirac} and investigated in detail by Pauli~\cite{Pauli}
and Sudarshan and
collaborators~\cite{Sudarshan,Sudarshan2,Sudarshan3}. We
investigate a method of field quantization motivated by Bender and
collaborators~\cite{Bender,Bender2,Bender3,Bender4,Brandt,Bender5}
that can implement a real energy spectrum for a pseudo-Hermitian
Hamiltonian, involving an indefinite metric in Hilbert space, and
a unitary $S$-matrix.

The stability of the vacuum is assured by postulating a positive
energy and a negative energy sector, which are identical copies of
the standard model of particles. The positive energy and negative
energy particles called positons and negatons, respectively, only
interact weakly through the gravitational field.  Moreover,
negative energy matter in the form of negaton particles does not
occur naturally in the universe. A negative energy ``shadow''
universe could exist separated from a positive energy universe
without being annihilated.

The positive and negative energy symmetry leads to a cancellation
of the zero-point vacuum energy in the presence of a gravitational
field and to the vanishing of the cosmological constant. Possible
corrections to the vanishing of the cosmological constant due to
interaction with the gravitational field can be small for a finite
quantum gravity.

A calculation of positon and negaton propagators leads to the
cancellation of quadratic divergences in the Higgs self-energy and
in two-point vacuum fluctuations.

\section{Indefinite Metric in Hilbert Space}

Dirac~\cite{Dirac,Pauli} generalized field quantization by
introducing an indefinite metric in the Hilbert space of the state
vectors. The normalization of a state vector $\Psi$ is normally
defined by
\begin{equation}
\label{positivenorm} {\cal N}_+=\int dq\Psi^*\Psi,
\end{equation}
where $\Psi^*$ is the complex conjugate of $\Psi$. The scalar
product of two complex state vectors $\Phi$ and $\Psi$ is given by
\begin{equation}
{\cal B}_+=\int dq\Phi^*\Psi.
\end{equation}
Instead, we consider the more general bilinear form
\begin{equation}
{\cal B}=\int dq\Phi^*\eta\Psi,
\end{equation}
in which the operator $\eta$ is an Hermitian operator to guarantee
real normalization values.

The expectation value of an observable ${\cal O}$ described by a
linear operator is now defined by
\begin{equation}
\langle{\cal O}\rangle=\int dq\Psi^*\eta{\cal O}\Psi.
\end{equation}
The generalization of the standard Hermitian conjugate operator
${\cal O}={\cal O}^\dagger$ is given by the adjoint operator
\begin{equation}
\tilde{\cal O}={\eta}^{-1}{\cal
O}^\dagger\eta^\dagger={\eta}^{-1}{\cal O}^\dagger\eta.
\end{equation}
All physical observables have to be self-adjoint, $\tilde{\cal
O}={\cal O}$, to guarantee that their expectation values are real.
In particular, the Hamiltonian operator $H$ has to be
self-adjoint, $\tilde H=H$, which has the consequence that
\begin{equation}
\frac{d}{dt}\int dq\Psi^*\eta\Psi=i\Psi^*\eta (\tilde H-H)\Psi=0,
\end{equation}
guaranteeing the conservation of the normalization with time.

A transformation of the Hermitian matrix $\eta$ to normal,
diagonal form can have the values $1$ or $-1$. The positive
definite form (\ref{positivenorm}) yields a unit matrix and
positive probabilities. However, in general positive eigenvalues
can have negative probabilities, i.e. one can introduce negative
probabilities that certain positive eigenvalues of an observable
are realized. We shall discuss the possibility of obtaining a
physical interpretation of quantum field theory in which the
$S$-matrix is unitary and only a positive energy spectrum is
observed when the pseudo-Hermitian Hamiltonian is quantized, and
is invariant under parity ${\cal P}$ and time reversal invariance
${\cal T}$.

\section{The Action and Field Quantization}

The action takes the form $(c=\hbar=1)$:
\begin{equation}
\label{gravaction} S=S_{\rm Grav}+S_M(\phi_+)+S_M(\phi_-),
\end{equation}
where
\begin{equation}
S_{\rm Grav}=\frac{1}{16\pi G}\int d^4x\sqrt{-g}[(R-2\Lambda_0)
-({\overline R}-2\overline\Lambda_0)].
\end{equation}
The $R$ denotes the normal Ricci scalar associated with positon
gravitons, while ${\overline R}$ is associated with negaton
gravitons. Moreover, $\Lambda_0$ and $\overline\Lambda_0$ denote
the ``bare'' cosmological constants corresponding to positon and
negaton gravitons, respectively. The $\phi_+$ and $\phi_-$ fields
denote positon and negaton matter fields, respectively.

We define an effective cosmological constant
\begin{equation}
\label{effectivelambda} \label{efflambda} \Lambda_{\rm
eff}=\Lambda_{0{\rm eff}}+\Lambda_{\rm vac},
\end{equation}
where $\Lambda_{0{\rm eff}}=\Lambda_0-\overline{\Lambda}_0$,
$\Lambda_{\rm vac}=8\pi G\rho_{\rm vac}$ and $\rho_{\rm vac}$
denotes the vacuum density.

We expand the metric tensor $g_{\mu\nu}$ about Minkowski flat
space
\begin{equation}
g_{\mu\nu}=\eta_{\mu\nu}+h_{\mu\nu}+O(h^2),
\end{equation}
where $\eta_{\mu\nu}={\rm diag}(1,-1,-1,-1)$. We will study the
lowest weak field approximation for which $\sqrt{-g}=1$.

Let us consider as a first simple case a real scalar field
$\phi(x)$ in the absence of interactions. The action is
\begin{equation}
\label{phiaction} S_\phi=\frac{1}{2}\int
d^4x\biggl[\partial_\mu\phi\partial^\mu\phi-\mu^2\phi^2\biggr],
\end{equation}
and $\phi$ satisfies the wave equation
\begin{equation}
\label{waveeq} (\partial^\mu\partial_\mu+\mu^2)\phi=0.
\end{equation}
As is well known, this equation has both positive and negative
energy solutions, as is the case with the Dirac
equation~\cite{Bjorken,Peskin}. We have
\begin{equation}
k_0\equiv\omega({\bf k})=\pm\sqrt{\vert{\bf k}\vert^2+\mu^2}.
\end{equation}
The equal time commutation relations for the field operator $\phi$
are
\begin{equation}
\label{comm1} [\phi({\bf x}),\phi({\bf x}')]=[\pi({\bf
x}),\pi({\bf x}')]=0,
\end{equation}
\begin{equation}
\label{comm2} [\phi({\bf x}),\pi({\bf x'})]=i\delta^3({\bf x}-{\bf
x}'),
\end{equation}
where $\pi={\dot\phi}$ is the conjugate momentum operator.

We shall retain both positive and negative energy solutions and
following Dirac~\cite{Dirac} and Pauli~\cite{Pauli}, decompose
$\phi$ into positive and negative energy parts:
\begin{equation}
 \phi(x)=A(x)+\tilde A(x).
 \end{equation}
 The quantization of $A(x)$ with $-k_0x_0$ in the phase factor
 occurs in the usual way, corresponding to positive energy
 particles, while the other part with $+k_0x_0$ in the phase
 factor is quantized such that it leads to negative energy
 particles. We have for $t=0$:
\begin{equation}
 \label{Aequation}
 A({\bf x})=\int \frac{d^3k}{((2\pi)^32k_0)^{1/2}}\{A_+(k)\exp[i({\bf k}\cdot{\bf x})]
+A_-(k)\exp[i(-{\bf k}\cdot{\bf x})]\},
 \end{equation}
 \begin{equation}
 \label{tildeAequation}
\tilde A(x) =\int \frac{d^3k}{((2\pi)^32k_0)^{1/2}}\{\tilde
A_+(k)\exp[i(-{\bf k}\cdot{\bf x})] +\tilde A_-(k)\exp[i({\bf
k}\cdot{\bf x})]\}.
 \end{equation}
The $\phi$ and $\pi$ operators are given by
 \begin{equation}
 \label{phi}
 \phi({\bf
 x})=\int\frac{d^3k}{[(2\pi)^32k_0]^{1/2}}\biggl\{(A_+(k)+{\tilde
 A}_-)\exp[i({\bf k}\cdot{\bf x})]
 $$ $$
+(A_-({\bf k})+{\tilde A}_+({\bf k}))\exp[i(-{\bf k}\cdot{\bf
x})]\biggr\},
 \end{equation}
 \begin{equation}
 \label{pi}
 \pi({\bf
 x})=\int\frac{d^3k}{[(2\pi)^3]}(-i)\sqrt{\frac{k_0}{2}}\biggl\{(A_+({\bf
 k})+{\tilde A}_-({\bf k}))\exp[i({\bf k}\cdot{\bf x})]
 $$ $$
-(A_-({\bf k})+{\tilde A}_+({\bf k}))\exp[i(-{\bf k}\cdot{\bf
x})]\biggr\}.
 \end{equation}

The equal time commutations relations for the $A$ operators assume
that all the $A_+({\bf k}),{\tilde A}_+$ operators commute with
the $A_-({\bf k}),{\tilde A}_-({\bf k})$ and the $A_+({\bf k})$
operators with the $A_+({\bf k})$ and the $A_-({\bf k})$ with the
$A_-({\bf k})$. Moreover, we have
\begin{equation}
[A_+({\bf k}),{\tilde A}_+({\bf k}')]=\delta^3({\bf k}-{\bf
k}'),\quad [A_-({\bf k}),{\tilde A}_-({\bf k}')]=-\delta^3({\bf
k}-{\bf k}').
\end{equation}
The equal time commutation relations (\ref{comm1}) and
(\ref{comm2}) follow from the commutation relations for the
$A_\pm$ operators.

The Hamiltonian takes the form
\begin{equation}
 H({\bf x})=\frac{1}{2}\int
 d^3x\biggl[({\bf\nabla}\phi({\bf x}))^2+(\partial_0\phi({\bf x}))^2+\mu^2\phi^2({\bf x})\biggr].
 \end{equation}
Substituting (\ref{phi}) and (\ref{pi}) into $H({\bf x})$, we get
\begin{equation}
\label{PTH} H({\bf x})=\frac{1}{(2\pi)^3}\int d^3kd^3k'\exp[i({\bf
k}+{\bf k}')\cdot{\bf x}]\biggl\{-\frac{\sqrt{\omega({\bf
k})\omega({\bf k'})}}{4}B({\bf k})+\frac{-{\bf k}\cdot{\bf
k}'+\mu^2}{4\sqrt{\omega({\bf k})\omega({\bf k'})}}C({\bf
k})\biggr\},
\end{equation}
where
\begin{equation}
B({\bf k})=[A_+({\bf k})+{\tilde A}_-({\bf k})-A_-(-{\bf k})
-{\tilde A}_+(-{\bf k})]
$$ $$
\times[A_+({\bf k}')+{\tilde A}_-({\bf k}')-A_-(-{\bf k}')-{\tilde
A}_+(-{\bf k}'],
\end{equation}
\begin{equation}
C({\bf k})=[A_+({\bf k})+{\tilde A}_-({\bf k})+A_-(-{\bf
k})+{\tilde A}_+(-{\bf k})]
$$ $$
\times[A_+({\bf k}')+{\tilde A}_-({\bf k}')+A_-(-{\bf k}')+{\tilde
A}_+(-{\bf k}')].
\end{equation}

By integrating $H({\bf x})$ we obtain
\begin{equation}
H\equiv \int d^3xH({\bf
x})=\frac{1}{(2\pi)^3}\int{d^3k}\omega({\bf k})[N_+({\bf
k})-N_-({\bf k})],
\end{equation}
where
\begin{equation}
N_+({\bf k})={\tilde A}_+({\bf k})A_+({\bf k}),\quad N_-({\bf
k})=-{\tilde A}_-({\bf k})A_-({\bf k}),
\end{equation}
denote the occupation numbers of the positons and negatons,
respectively. The field momentum is given by
\begin{equation}
{\bf P}=\frac{1}{(2\pi)^3}\int{d^3k}{\bf k}[N_+({\bf k})-N_-({\bf
k})].
\end{equation}
We see that the {\it zero-point vacuum energy contributions
corresponding to the infinite c-number $\delta(0)$ have
cancelled}. The Fock vacuum is the state which satisfies
\begin{equation}
A_+({\bf k})\vert 0\rangle=0,\quad A_-({\bf k})\vert 0\rangle=0,
\end{equation}
corresponding to the eigenvalue $E_{\rm vac}=0$.

In standard second quantized quantum field theory with only
positon scalar fields, we have
\begin{equation}
E=\sum_kk_0\biggl[\frac{1}{2}+N_{+k}\biggr].
\end{equation}
For the vacuum (ground state), $N_{+k}=0$, and we obtain the
zero-point vacuum energy
\begin{equation}
\label{zeropointenergy} E_{\rm vac}\equiv
E_0=\frac{1}{2}\sum_kk_0=\frac{1}{2(2\pi)^3}\int d^3k\sqrt{\vert
{\bf k}\vert^2+\mu^2}.
\end{equation}
The zero-point vacuum energy $E_0$ diverges quartically and is the
root of the cosmological constant problem in the presence of a
gravitational field, since graviton loops can couple to the vacuum
energy ``bubble'' graphs which cannot be time-ordered away, i.e.
we cannot simply shift the infinite constant vacuum energy, $E_0$,
such that only $E'=E-E_0$ is observed.

An alternative quantization procedure consists of defining besides
the field $\phi(x)$ another scalar field $\chi(x)$, the adjoint of
which is $\tilde\chi(x)=-\chi(x)$~\cite{Dirac,Pauli}. Then we have
\begin{equation}
\chi(x)=\frac{1}{\sqrt{2}}[A(x)-\tilde A(x)],
\end{equation}
with the Fourier decomposition
\begin{equation}
 \chi(x)=V^{-1/2}\sum_k(2k_0)^{-1/2}\{\tilde\chi(k)\exp[i({\bf k}\cdot{\bf x}-k_0x_0)]
 -\chi(k)\exp[i(-{\bf k}\cdot{\bf x}+k_0x_0)]\}.
 \end{equation}
 The $\chi$ field is quantized according to
 \begin{equation}
 [\chi(k),\tilde\chi(k)]=-1,
 \end{equation}
 which gives
 \begin{equation}
 \tilde\chi(k)\chi(k)=-N_{\chi}(k).
 \end{equation}
 The Hamiltonian is now given by
 \begin{equation}
 H=\frac{1}{2}\int
 d^3x\biggl[({\bf\nabla}\phi)^2+(\partial_0\phi)^2+\mu^2\phi^2
 -(\vec\nabla\chi)^2-(\partial_0\chi)^2-\mu^2\chi^2\biggr].
 \end{equation}
 This leads to the energy
 \begin{equation}
 E=\sum_kk_0[N_\phi-N_\chi].
 \end{equation}
 As before, the positive and negative energy symmetry of the vacuum state
 leads to the cancellation of the zero-point vacuum energy, $E_0$.
 We have
 \begin{equation}
 \phi(k)=\frac{1}{\sqrt{2}}[A_+(k)+\tilde A_-(k)],\quad
 \tilde\phi(k)=\frac{1}{\sqrt{2}}[\tilde A_+(k)+A_-(k)],
 \end{equation}
 \begin{equation}
 \chi(k)=\frac{1}{\sqrt{2}}[\tilde A_+(k)-A_-(k)],\quad
 \tilde\chi(k)=\frac{1}{\sqrt{2}}[A_+(k)-\tilde {A}_-(k)].
 \end{equation}

A spinor field $\psi$ satisfies in the presence of an
electromagnetic interaction the Dirac equation
\begin{equation}
[\gamma^\mu (p_\mu-eA_\mu)-m]\psi=0,
\end{equation}
and its charge conjugate equation
\begin{equation}
[\gamma^\mu(p_\mu+eA_{c\mu})-m]\psi_c=0.
\end{equation}
These equations yield both positive and negative energy solutions
of the Dirac equation. The two spinor fields $\psi$ and $\psi_c$
and the two photon fields $A_\mu$ and $A_{c\mu}$ are associated
with positive and negative energy fermions and neutral gauge
fields, respectively. We introduce the positive and negative
energy spinor field
\begin{equation}
\psi(x)=\Psi(x)+{\tilde\Psi}(x),
\end{equation}
and for the photon field we define
\begin{equation}
A_\mu(x)=U_\mu(x)+{\tilde U} _\mu(x).
\end{equation}
For the charge conjugation transformation we have
\begin{equation}
\psi_c=C\gamma^0\psi^*=C{\overline\psi}^T,
\end{equation}
where $C$ is the charge conjugation matrix which satisfies
\begin{equation}
C^{-1}\gamma^\mu C=-\gamma^{\mu T},
\end{equation}
and $\gamma^{\mu T}$ denotes the transpose of the Dirac $\gamma$
matrix. A similar transformation exists for the gauge field
$A_\mu$ which transforms it into the anti-particle gauge field.

We have the spinor wave expansion
\begin{equation}
\psi(x)=\frac{1}{(2\pi)^{3/2}}\int
d^3p\sqrt{\frac{m}{E_p}}\{b_{+r}(p)u_{+r}(p)\exp(-ip\cdot x)
+{\tilde d}_{+r}(p)v_{+r}(p)\exp(ip\cdot x)
$$ $$
+ b_{-r}(p)u_{-r}(p)\exp(-ip\cdot x) + d_{-r}(
p)v_{-r}(p)v_{-r}(p)\exp(-ip\cdot x)\},
\end{equation}
where $E_p=\sqrt{\vert{\bf p}\vert^2+m^2}$. The conjugate momentum
is $\pi(x)=i{\tilde\psi}(x)$ and the $\psi(x)$ satisfy the
anti-commutation relation
\begin{equation}
\{\psi({\bf x},t),{\tilde\psi}({\bf x}',t)\}=\delta^3({\bf x}-{\bf
x}').
\end{equation}

The Hamiltonian is given by
\begin{equation}
H=\int d^3x[{\tilde\psi}(x)(-i({\bf\alpha}\cdot{\bf\Delta}+\beta
m)\psi]=i\int d^3x{\tilde\psi}(x)\partial_0\psi(x),
\end{equation}
where $\alpha$ and $\beta$ are Dirac matrices. The
anti-commutation relations for the $b$ and $d$ operators are of
the form
\begin{equation}
\{b_{+r}(p),{\tilde b}_{+r'}(p)\}=\delta_{rr'}\delta^3({\bf
p}-{\bf p}),\quad \{d_{+r}(p),{\tilde
d}_{+r'}(p)\}=\delta_{rr'}\delta^3({\bf p}-{\bf p}),
$$ $$
\{b_{-r}(p),{\tilde b}_{-r'}(p)\}=-\delta_{rr'}\delta^3({\bf
p}-{\bf p}),\quad \{d_{-r}(p),{\tilde
d}_{-r'}(p)\}=-\delta_{rr'}\delta^3({\bf p}-{\bf p}).
\end{equation}
All the other anti-commutation relations of the $b_+,d_+,b_-,d_-$
vanish. We obtain for the energy:
\begin{equation}
\label{fermionenergy} E=\int d^3pE_p[N^{(+)}_+({\bf
p})-N^{(-)}_+({\bf p}) -N^{(+)}_-({\bf p})+N^{(-)}_-({\bf p})].
\end{equation}
The occupation numbers are defined by
\begin{equation}
N^{(+)}_+({\bf p})={\tilde b}_+({\bf p})b_+({\bf p}), \quad
N^{(-)}_+({\bf p})={\tilde d}_+({\bf p})d_+({\bf p}),
$$ $$
N^{(+)}_-({\bf p})=-{\tilde b}_-({\bf p})b_-({\bf p}),\quad
N^{(-)}_-({\bf p})=-{\tilde d}_-({\bf p})d_-({\bf p}).
\end{equation}
The vacuum state is
\begin{equation}
N^{(+)}_+({\bf p})\vert 0\rangle=0,\quad N^{(-)}_+({\bf p})\vert
0\rangle=0,\quad N^{(+)}_-({\bf p})\vert 0\rangle=0,\quad
N^{(-)}_-({\bf p})\vert 0\rangle=0.
\end{equation}

We see from (\ref{fermionenergy}) that the zero-point vacuum
energy for fermions has cancelled. In the standard treatment of
the quantization of Dirac spinors, a normal ordering of the
operators results in the cancellation of the infinite zero-point
vacuum energy. However, this entails subtracting a quartically
divergent (infinite) constant from the actual energy. The
cancellation of the vacuum energy, due to the positive and
negative energy symmetry of the vacuum, does not invoke unphysical
infinite constants. Moreover, in the presence of a gravitational
field the normal ordering of the fermion operators is not valid.

The normalization of the state vector $\Psi$ is determined
by~\cite{Pauli}:
\begin{equation}
{\cal N}=\sum_{N_+(k),N_-(k)}(-1)^{\sum
N_-(k)}\Psi^*(...N_+(k)...,...N_-(k)...)
$$ $$
\times\Psi(...N_+(k)..., ...N_-(k)...)={\rm const.}
\end{equation}
This demonstrates that ``negative probability'' states will exist
with an odd number of particles in states with negative energy. We
shall see in the following section, how we can quantize the field
in the presence of interactions and avoid a catastrophic
instability due to these negative probabilities and negative
energy particles. In Section 5, we shall investigate how we can
formulate the quantum field theory, so that we obtain a real and
positive energy spectrum and a unitary $S$-matrix.

It can be shown that an equivalent quantization procedure for
complex charged scalar fields, neutral vector gauge fields
$A_\mu$, spin-2 graviton fields, can be derived that leads to the
cancellation of the zero-point vacuum energy in the presence of a
gravitational field. This is again due to the positive-negative
energy symmetry of the vacuum state $\vert 0\rangle$ for these
fields.

\section{Positive and Negative Energy Mirror Sectors}

In standard second quantized field theory and in the current
interpretation of the standard model, the creation and
annihilation operators for a negative energy particle are
interpreted as the annihilation and creation of a positive energy
particle with the opposite charge $-\vert e\vert$ corresponding to
an antiparticle. The negative energy particle vacuum is empty as
is the vacuum of positive energy particles.

We must now assure that the visible positons are stable against
decay into negatons and that the annihilation of positons and
negatons does not destablize the vacuum\footnote{In previous
work~\cite{Moffat}, we postulated that the vacuum was fully
occupied by negative energy fermions and bosons. The negative
energy bosons were required to satisfy parastatistics in an
attempt to obtain a Pauli exclusion principle for the sea of
negative energy bosons. However, any number of parabosons can
occupy an antisymmetric quantum state, excluding the possible
existence of a paraboson exclusion principle~\cite{Greenberg}.}.
The problem of the stability of matter when negative energy
particles are included has been investigated~\cite{Carroll,
Cline}. When an ordinary positive energy positon collides with a
negaton, the positon energy can {\it increase} due to the
compensating negative energy of the negaton. Since there is an
infinite amount of phase space available, the rate of decay is
infinite when arbitrarily high momenta are included; the negaton
particles can have arbitrarily large negative energies.

We postulate the existence of a visible positon matter sector and
a negative energy negaton matter ``shadow'' sector, which are
identical copies of the standard model of particles. These two
sectors only couple weakly through gravity

Let us denote positon particles with positive energy by $\phi_+$
and negatons with negative energy by $\phi_-$. Consider the
Feynman tree-graph decay of a positon particle $\phi_{1+}$ into a
positon $\phi_{2+}$ plus a negaton $\phi_-$:
\begin{equation}
\phi_{1+}\rightarrow \phi_{2+}+\phi_-.
\end{equation}
Neither this reaction not the reaction
\begin{equation}
\phi_{1+}+\phi_-\rightarrow \phi_{2+},
\end{equation}
can occur through standard model couplings. Similarly, the process
\begin{equation}
e^{(-)}_+ \rightarrow \mu^{(-)}_+
+\nu_{e+}+{\bar\nu}_{\mu+}+\phi_-^{(0)},
\end{equation}
is not allowed.  Here, $e^{(-)}_+$, $e^{(+)}_-$, $\mu^{(-)}_+$,
${\bar\nu}_{\mu+}$, $\phi_-^{(0)}$ denote the positive energy
electron, positron, muon, and neutral negaton, respectively,
Moreover, the energy conserving annihilation process
\begin{equation}
e^{(-)}_+ + e^{(+)}_+\rightarrow \gamma_-+\gamma_-,
\end{equation}
where $\gamma_-$ denotes the negaton photon. Similar reactions of
positon quarks into negaton gluons, for example, are also
forbidden.

We also postulate that negative energy matter is not found
naturally on Earth and in the universe, except in vanishingly
small quantities. This is similar to positive energy antimatter,
which is also not found naturally in the universe and very briefly
in vanishingly small amounts as the results of radioactive decay
or cosmic rays. This avoids the catastrophic annihilation of
positon and negaton matter and guarantees the meta-stability of
the vacuum. The matter-antimatter symmetry was broken in the very
early universe. In a similar way, the positon-negaton symmetry was
also broken in the early universe, avoiding the instability of
Minkowski spacetime. As in the breaking of matter-antimatter
symmetry in the early universe, the breaking of positive-negative
energy symmetry requires an explanation.

We must consider the possible decay of negative energy negaton
particles. Consider the tree-graph decays of negatons:
\begin{equation}
\phi_-\rightarrow \phi_{1+}+\phi_{2+}.
\end{equation}
This is forbidden since
\begin{equation}
\vert 0\rangle\rightarrow \phi_{-}+\phi_{1+}+\phi_{2+}
\end{equation}
is not allowed. But the decay into one negaton and
one positon is allowed
\begin{equation}
\phi_{1-}\rightarrow \phi_{2-}+\phi_+,
\end{equation}
provided that
\begin{equation}
\phi_{2-}\rightarrow \phi_{1-}+\phi_+.
\end{equation}
This reaction requires that $m_2
> m_1+m_\phi$. To avoid any catastrophic instability
of negaton reactions, we postulate that the interaction of negaton
particles with one another is weak, $\alpha_{\rm neg} \ll
\alpha_w$, where $\alpha_w \sim 10^{-6}$ denotes the electroweak
standard model coupling constant.

Positon and negaton gravitons can interact with the visible
positon matter and the shadow negaton matter. The negaton decay
channel involving the smallest number of positons and a positon
graviton is
\begin{equation}
\phi_-\rightarrow g_+ +\phi_{1-}+\phi_{2-},
\end{equation}
where $g$ denotes a graviton. An effective field theory with a
cutoff $\Lambda_{\rm cut}$ can lead to a sufficiently stable
vacuum, provided that higher derivative couplings are suppressed
by the cutoff~\cite{Carroll,Cline}.

These postulates lead to a meta-stable vacuum that has a life-time
greater than the age of the universe, i.e., a lifetime greater
than the Hubble time, $H_0^{-1}\sim 10^{60}M_P^{-1}$, where $H_0$
and $M_P$ denote the Hubble constant and the Planck mass,
respectively.

In contrast to supersymmetry, we do not postulate new species of
particles such as supersymmetry partners with differing spins and
satisfying opposite particle statistics. The dual energy symmetry
holds for all known species of particles and corresponds to a
doubling of the degrees of freedom of know standard model
particles.

\section{Pseudo-Hermitian Hamiltonian and Unitarity}

In standard quantum field theory the Hamiltonian is Hermitian,
$H^\dagger=H$, and we are guaranteed that the energy spectrum is
real and positive and that the time evolution of the operator
$U=\exp(itH)$ is unitary and probabilities are positive and
preserved for particle transitions. However, in recent years there
has been a growth of activity in studying quantum theories with
pseudo-Hermitian Hamiltonians, which satisfy the generalized
property of adjointness, $\tilde H=\eta^{-1}H^\dagger\eta$=H,
associated with an indefinite metric in Hilbert
space~\cite{Bender,Bender2,Bender3,Bender4,Brandt,Bender5}.

Spectral reality and unitarity can in special circumstances follow
from a symmetry property of the Hamiltonian in terms of the
symmetry under the operation of ${\cal P}{\cal T}$, where ${\cal
P}$ is a linear operator represented by parity reflection, while
${\cal T}$ is an anti-linear operator represented by time
reversal. If a Hamiltonian has an unbroken ${\cal P}{\cal T}$
symmetry, then the energy levels can in special cases be real and
the theory can be unitary and free of ``ghosts''. The operation of
${\cal P}$ leads to ${\bf x}\rightarrow -{\bf x}$, while the
anti-linear operation of ${\cal T}$ leads to $i\rightarrow -i$. It
follows that under the operation of ${\cal P}{\cal T}$ the
Hamiltonian $H$ in (\ref{PTH}) is invariant under the ${\cal
P}{\cal T}$ transformation, which is necessary but not sufficient
to assure the reality of the energy eigenvalues.

The proof of unitarity follows from the construction of a linear
operator ${\cal C}$. This operator is used to define the inner
product of state vectors in Hilbert space:
\begin{equation}
\label{innerprod} \langle\Psi\vert\Phi\rangle=\Psi^{{\cal C}{\cal
P}{\cal T}}\cdot\Phi.
\end{equation}
Under general conditions, it can be shown that a necessary and
sufficient condition for the existence of the inner product
(\ref{innerprod}) is the reality of the energy spectrum of $H$.
With respect to this inner product, the time evolution of the
quantum theory is unitary. In quantum mechanics and in quantum
field theory, the operator ${\cal C}$ has the general form
\begin{equation}
{\cal C}=\exp(Q){\cal P},
\end{equation}
where $Q$ is a function of the dynamical field theory variables.
The form of ${\cal C}$ must be determined by solving for the
function $Q$ in terms of chosen field variables and field
equations. The form of ${\cal C}$ has been calculated for several
simple field theories, e.g. $\phi^3$ theory and also in massless
quantum electrodynamics with a pseudo-Hermitian Hamiltonian. The
solution for ${\cal C}$ satisfies
\begin{equation}
{\cal C}^2=1,\quad [{\cal C},{\cal P}{\cal T}]=0,\quad [{\cal
C},H]=0.
\end{equation}
We shall not attempt to determine a specific generalized charge
conjugation operator ${\cal C}$ in the present work.

It has also been shown that a special form of ${\cal P}$ leads to
a Lorentz invariant scalar expression for the operator ${\cal
C}$~\cite{Brandt}.

\section{Propagators and Vacuum Fluctuations}

We can evaluate the field commutator for $\phi(x)$:
\begin{equation}
[\phi(x),\phi(x')]=\frac{1}{(2\pi)^3}\int\frac{d^3kd^3k'}{\sqrt{2\omega_k2\omega_{k'}}}
\biggl\{[A_+(k),{\tilde A}_+(k')]\exp[-ik{\cdot x}+ik'{\cdot{x}'}]
$$ $$
+[{\tilde A}_+(k),A_+(k')]\exp[ik\cdot x-ik'\cdot{x}']
$$ $$
+ [A_-(k),{\tilde A}_-(k')]\exp[ik{\cdot
x}-ik'{\cdot{x}'}]+[{\tilde A}_-(k),A_-(k')]\exp[-ik\cdot
x+ik'\cdot{x}']
$$ $$
=\frac{2}{(2\pi)^3}\int\frac{d^3k}{2\omega_k}\{\exp[-ik\cdot(x-x')]
-\exp[ik\cdot(x-x')]\}.
\end{equation}
We obtain the result
\begin{equation}
\frac{1}{2}[\phi(x),\phi(x')]=i\Delta(x-x').
\end{equation}
The equal time commutator of the $\phi$ fields vanishes for
space-like separation $(x-x')^2 < 0$ in agreement with Lorentz
invariance and microscopic causality.

Let us define for charged scalar particles
\begin{equation}
\phi(x)=\frac{1}{2}[\phi_1(x)+i\phi_2(x)],
\end{equation}
where
\begin{equation}
(\partial_\mu\partial^\mu+\mu^2)\phi_1=0,\quad
(\partial_\mu\partial^\mu+\mu^2)\phi_2=0.
\end{equation}
Then, the commutator for the charged scalar fields is
\begin{equation}
\frac{1}{2}[\phi_i(x),\phi_j(x)]=i\delta_{ij}\Delta(x-x').
\end{equation}
The commutators of $\phi_i$ with $\phi_j$ and $\phi^*_i$ with
$\phi^*_j$ vanish.

We define the Feynman propagator for the positons and negatons:
\begin{equation}
i\Delta_F(x'-x)=\langle 0\vert\phi(x')\phi^*(x)\vert 0\rangle
\theta(t'-t)+\langle 0\vert\phi^*(x)\phi(x')\vert
0\rangle\theta(t-t').
\end{equation}
We obtain the modified Feynman propagator:
\begin{equation}
\label{Feynman} i\Delta_F(x'-x) =\frac{i}{(2\pi)^4}\int
d^4k\biggl\{\exp[-ik\cdot(x-x')]\biggl[\frac{1}{k^2-\mu^2+i\epsilon}
-\frac{1}{k^2-\mu^2-i\epsilon}\biggr]\biggr\}.
\end{equation}
We have the result that
\begin{equation}
(\partial_\mu\partial^\mu+\mu^2)\Delta^+_F(x'-x)=\delta^4(x'-x),
\end{equation}
\begin{equation}
(\partial_\mu\partial^\mu+\mu^2)\Delta^-_F(x'-x)=\delta^4(x'-x),
\end{equation}
where
\begin{equation}
\Delta_F(x'-x)=\frac{1}{2}[\Delta^+_F(x'-x)+\Delta^-_F(x'-x)],
\end{equation}
\begin{equation}
\Delta^+_F(x'-x)=\frac{i}{2(2\pi)^4}\int
d^4k\frac{\exp[-ik\cdot(x-x')]}{k^2-\mu^2+i\epsilon},
\end{equation}
\begin{equation}
 \Delta^-_F(x'-x)=
 -\frac{i}{2(2\pi)^4}\int
d^4k\frac{\exp[-ik\cdot{(x-x')}]}{k^2-\mu^2-i\epsilon}.
\end{equation}
We also find that
\begin{equation}
i\Delta_F(0)=\frac{i}{2(2\pi)^4}\int
d^4k\biggl[\frac{1}{k^2-\mu^2+i\epsilon}
-\frac{1}{k^2-\mu^2-i\epsilon}\biggr]=\frac{1}{16\pi^3}\int
d^4k\delta(k^2-\mu^2),
\end{equation}
where we have used the identity
\begin{equation}
\label{delta}
\frac{1}{k^2-\mu^2+i\epsilon}-\frac{1}{k^2-\mu^2-i\epsilon}=-2i\pi\delta(k^2-\mu^2).
\end{equation}
We observe that $i\Delta_F(0)$ has a finite value in contrast to
the usual result for positons that $i\Delta_F(0)$ diverges
quadratically.

A calculation of the two-point vacuum fluctuations yields
\begin{equation}
\langle 0\vert\phi(x)\phi(x')\vert 0\rangle
=\frac{1}{(2\pi)^3}\int
\frac{d^3k}{2k_0}\biggl(\exp[-ik\cdot(x-x')]-\exp[ik\cdot(x-x')]\biggr).
\end{equation}
We find that for $x=x'$:
\begin{equation}
\langle 0\vert\phi^2(0)\vert 0\rangle=\Delta(0)=0.
\end{equation}
Thus, due to the positive and negative energy symmetry, we avoid
the quadratic divergence of the two-point vacuum fluctuations in
standard second quantization for positons.

\section{Resolution of the Cosmological Constant and Higgs
Hierarchy Problems}

We shall postulate that
\begin{equation}
\label{barelambda} \Lambda_{0{\rm
eff}}=\Lambda_0-{\overline\Lambda}_0=0.
\end{equation}
The vanishing of the zero-point vacuum energy in our quantum field
theory, including higher-order graviton tree-graph couplings and
loops, then assures that (\ref{barelambda}) is protected against
all higher order radiative vacuum corrections.

If we assume that a spontaneous symmetry breaking of the positive
and negative energy symmetry of the vacuum occurs, then this will
create a small `` observed'', effective cosmological constant
$\Lambda_{\rm eff}/8\pi G\sim (2\times 10^{-3}\,eV)^4$, needed to
provide a cosmological constant explanation of the accelerating
expansion of the universe~\cite{Perlmutter,Riess,Spergel}.
However, it is possible that the accelerating expansion of the
universe can be explained by a late-time inhomogeneous
cosmological model~\cite{Moffat2,Moffat3,Moffat4}, in which the
cosmological constant $\Lambda_{\rm eff}=0$ and there is no need
for a negative pressure ``dark energy''. Moreover, we also have to
account for small higher-order gravitational corrections.

Let us consider the scalar field theory with the interaction term:
\begin{equation}
S_I=\frac{\lambda}{4!}\int d^4x\phi^4(x).
\end{equation}
We shall use the propagator (\ref{Feynman}) in momentum space to
calculate the self-energy of the scalar field:
\begin{equation}
\Sigma(p^2)\equiv \Sigma(0)=\frac{\lambda}{2(2\pi)^4}\int
d^4k\biggl(\frac{1}{k^2-m^2+i\epsilon}
-\frac{1}{k^2-m^2-i\epsilon}\biggr).
\end{equation}

We now obtain the result by using (\ref{delta}):
\begin{equation}
\Sigma(0)=\frac{-i\pi\lambda}{(2\pi)^4}\int d^4k\delta(k^2-m^2).
\end{equation}
Transforming to Euclidean momentum space we have: $k^0=ik^{0'}$,
$k^{'2}={\bf k}^{'2}+k^{'02}$, $d^4k=id^4k'$ and $\int
d^4k'=\pi^2\int dk^{'2}k^{'2}$ which yields
\begin{equation}
\Sigma(0)=\frac{\lambda m^2}{16\pi}.
\end{equation}
We see that $\Sigma(0)$ is finite and we avoid the quadratic
divergence of the scalar field self-energy in standard QFT. This
resolve the Higgs hierarchy problem in the standard model.

A Casimir vacuum energy has been experimentally observed. In our
quantum field theory, the vanishing of the zero-point vacuum
energy is only valid in the absence of material boundary
conditions as are necessary for the Casimir effect~\cite{Jaffe}.
When material boundary conditions such as the parallel metal
plates required to perform the Casimir experiments are imposed,
then we can no longer demand that the positive and negative energy
symmetry of the vacuum state is preserved; the breaking of this
symmetry of the vacuum will produce a non-vanishing zero-point
energy effect.

Jaffe~\cite{Jaffe} points out that the Casimir effect gives no
more or less evidence for the ``reality'' of the vacuum
fluctuation energy of quantum fields than any other one-loop
effect in quantum electrodynamics, e.g. the vacuum polarization
effect associated with charges and currents in atomic physics.
Like all other observable effects in quantum electrodynamics, the
Casimir effect vanishes as the fine structure constant $\alpha$
goes to zero.

\section{Conclusions}

We have formulated a quantum field theory based on an indefinite
metric in Hilbert space with a generalization of the Hermitian
Hamiltonian operator $H=H^\dagger$ to an adjoint operator $\tilde
H=\eta^{-1}H^\dagger\eta$ and we have ${\tilde H}=H$. The
quantization of fields in the presence of gravity is performed
with a positive and negative energy particle interpretation, which
leads to the cancellation of the zero-point vacuum energy due to
the positive and negative dual energy symmetry of the vacuum. We
have
\begin{equation}
H=H_+ + H_-,
\end{equation}
where
\begin{equation}
\quad H_+\vert 0\rangle=E_{\rm vac}\vert 0\rangle,\quad H_-\vert
0\rangle = -E_{\rm vac}\vert 0\rangle,
\end{equation}
and
\begin{equation}
\label{vacH} \langle 0\vert H\vert 0\rangle=0.
\end{equation}
We postulate that the effective, classical  ``bare'' cosmological
constant $\Lambda_{0{\rm eff}}=0$. The condition (\ref{vacH})
leads to a protection of the vanishing of the effective
cosmological constant, $\Lambda_{\rm eff}$, from gravitational and
external field quantum corrections. Any possible higher-order
gravitational corrections will be expected to be small in a finite
quantum gravity theory. This can be interpreted as a resolution of
the cosmological constant problem. The scalar field
self-interaction is finite and can resolve the standard model
Higgs hierarchy problem.

The stability of the vacuum and positive energy standard model
particles is assured by postulating two identical standard model
particle sectors, a visible positon matter sector and a shadow
negaton matter sector, which interact only weakly through
gravitational interactions or the curved geometry of spacetime. A
similar scenario has been proposed by Kaplan and Sundrum
~\cite{Sundrum}. Possible observational consequences of this
gravitational interaction between the two sectors for the early
universe will be considered in a future publication.

The indefinite Hilbert space state vector metric can generate
negative probabilities for the transitions of particles and
violate the unitarity of the $S$-matrix. To guarantee positive
probabilities and the unitarity of transition and scattering
amplitudes, we incorporate the ${\cal P}{\cal T}$ operation on
field operators and the action. The generalized charge conjugation
operator ${\cal C}$, introduced by Bender and
collaborators~\cite{Bender,Bender2,Bender3,Bender4,Bender5} is
invoked to affirm that the energy spectrum for gravity and the
standard model particle theory is positive, and assure that
probabilities are positive and conserved and that the $S$-matrix
is unitary. Further investigation of the properties of the
operator ${\cal C}$ that will guarantee a physical energy spectrum
and unitary S-matrix for the standard model and gravity will be
presented elsewhere.

\vskip 0.2 true in{\bf Acknowledgment} \vskip 0.2 true in

This research was supported by the Natural Sciences and
Engineering Research Council of Canada. I thank Joel Brownstein
and Martin Green for helpful and stimulating discussions.

\end{document}